\documentclass{article}

\usepackage{amsmath}
\usepackage{graphicx}
\usepackage[colorlinks=true, allcolors=blue]{hyperref}
\usepackage[english]{babel}
\usepackage[letterpaper,top=2cm,bottom=2cm,left=3cm,right=3cm,marginparwidth=1.75cm]{geometry}
\usepackage[nolist]{acronym}
\usepackage{cleveref}
\usepackage{tabularx}
\usepackage[table]{xcolor}
\usepackage{adjustbox}
\usepackage{booktabs}
\usepackage{siunitx}
\usepackage{subcaption}
\usepackage{makecell}
\usepackage{authblk}
\usepackage{tikz,xcolor,hyperref}
\usepackage{orcidlink}
\definecolor{lime}{HTML}{A6CE39}
\definecolor{orangescid}{HTML}{F7AF72}
\DeclareRobustCommand{\orcidicon}{
	\begin{tikzpicture}
	\draw[lime, fill=lime] (0,0) 
	circle [radius=0.16] 
	node[white] {{\fontfamily{qag}\selectfont \tiny ID}};
	\draw[white, fill=white] (-0.0625,0.095) 
	circle [radius=0.007];
	\end{tikzpicture}
	\hspace{-2mm}
}
\foreach \x in {A, ..., Z}{\expandafter\xdef\csname orcid\x\endcsname{\noexpand\href{https://orcid.org/\csname orcidauthor\x\endcsname}
			{\noexpand\orcidicon}}
}

\title{Real-world energy data of 200 feeders from low-voltage grids with metadata in Germany over two years}

\newcommand{\corrauth}{\thanks{Corresponding author: Manuel Treutlein (m.treutlein@netze-bw.de, manuel.treutlein@partner.kit.edu). Further e-mail addresses: p.bothe@netze-bw.de, marc.schmidt@netze-bw.de, r.hahn@netze-bw.de, o.neumann@netze-bw.de, ralf.mikut@kit.edu, veit.hagenmeyer@kit.edu}}
\author[1,2]{Manuel Treutlein \orcidlink{0009-0006-1071-341X}\corrauth}

\author[2]{Pascal Bothe \orcidlink{0000-0003-0262-1102}}

\author[2]{Marc Schmidt \orcidlink{0000-0002-9127-1708}}

\author[2]{Roman Hahn \orcidlink{0009-0002-2115-3711}}

\author[2]{Oliver Neumann \orcidlink{0000-0003-4438-300X}}

\author[1]{Ralf Mikut \orcidlink{0000-0001-9100-5496}}

\author[1]{Veit Hagenmeyer \orcidlink{0000-0002-3572-9083}}

\affil[1]{Institute for Automation and Applied Informatics (IAI), Karlsruhe Institute of Technology (KIT), Germany}

\affil[2]{AI \& Analytics, Netze BW GmbH, Germany}

\begin{document}

\begin{acronym}
    \acro{LCT}[LCT]{low-carbon technology}
    \acro{LV}[LV]{low-voltage}
    \acro{MV}[MV]{medium-voltage}
    \acro{DSO}[DSO]{distribution system operator}
    \acro{PF}[PF]{power flow}
    \acro{SE}[SE]{state estimation}
    \acro{ML}[ML]{machine learning}
    \acro{PV}[PV]{photovoltaic}
    \acro{EV}[EV]{electric vehicle}
    \acro{CHP}[CHP]{combined heat and power}
    \acro{NILM}[NILM]{non-intrusive load monitoring}
    \acro{DOI}[DOI]{digital object identifier}
    \acro{RMS}[RMS]{root mean square}
    \acro{PEN}[PEN]{protective earth neutral}
    \acro{NWP}[NWP]{numerical weather prediction}
\end{acronym}

\maketitle

\begin{abstract}
The last mile of the distribution grid is crucial for a successful energy transition, as more low-carbon technology like photovoltaic systems, heat pumps, and electric vehicle chargers connect to the low-voltage grid. Despite considerable challenges in operation and planning, researchers often lack access to suitable low-voltage grid data. To address this, we present the FeederBW dataset with data recorded by the German distribution system operator Netze BW. It offers real-world energy data from 200 low-voltage feeders over two years (2023-2025) with weather information and detailed metadata, including changes in low-carbon technology installations. The dataset includes feeder-specific details such as the number of housing units, installed power of low-carbon technology, and aggregated industrial energy data. Furthermore, high photovoltaic feed-in and one-minute temporal resolution makes the dataset unique. FeederBW supports various applications, including machine learning for load forecasting, conducting non-intrusive load monitoring, generating synthetic data, and analyzing the interplay between weather, feeder measurements, and metadata. The dataset reveals insightful patterns and clearly reflects the growing impact of low-carbon technology on low-voltage grids.
\end{abstract}

\maketitle

\section*{Background \& Summary}\label{sec1}
The \ac{LV} grid is pivotal in facilitating the energy transition towards renewable energy sources and increased electrification. Integrating essential \ac{LCT} such as heat pumps, \acp{EV}, and \ac{PV} systems into the  \ac{LV} grid in substantial numbers has become an indispensable task. This shift requires significant modifications in the operation and planning of \ac{LV} grids, which were traditionally designed for fossil-fuel-based systems. For example, planners and operators have to deal with much more data from decentralized generation and consumption \cite{rolnickTacklingClimateChange2023}. Consequently, a large and active research community has been focusing on developing innovative approaches to make this transition economically viable. A critical component of this research is the availability of realistic data, a need that is further amplified by advancements in machine learning and artificial intelligence which requires large amounts of data \cite{habenReviewLowVoltage2021}.

The energy transition in the \ac{LV} grid presents diverse challenges from the perspective of \acp{DSO}. Grid planners are tasked with dimensioning existing and new \ac{LV} grids to ensure adequate capacity. To achieve this, they employ \ac{PF} calculations, coincidence factors, or load and generation profiles. However, the effectiveness of these methods hinges on the availability of realistic data, which should ideally include power measurements of \ac{LCT} for which high future penetration rates are anticipated. Regarding the operation of \ac{LV} grids, it is increasingly critical for operators to have grid \acp{SE} based on real measurements and models. This enables continuous monitoring but also serves as a prerequisite for controlling load and generation to prevent overloads or voltage band violations in cables, overhead lines, or transformers.

Furthermore, real measurement data is essential for training machine learning models that support \ac{DSO} processes. Nevertheless, many \ac{LV} grids lack sufficient measurement devices, because they are not yet installed or too expensive. In addition, future measurements are inherently unavailable. Consequently, models based on real-time data and realistic historical data are crucial. For instance, these models can produce load forecasts for \ac{LV} feeders for the following day or help addressing the gaps left by insufficient measurement infrastructure.

Several review papers focusing on \ac{LV} grids and smart grids, which include overviews of public datasets, have been published in recent years \cite{habenReviewLowVoltage2021, altamimiSmartGridPublic2024, elaheReviewLoadData2021}. However, authors across these review papers consistently highlight the lack of sufficient public datasets. Research on non-public datasets is frequently observed in \ac{LV} load forecasting \cite{habenReviewLowVoltage2021}, leading to non-reproducible research and contradicting good scientific practice. A major issue is that many power companies refuse to provide public datasets due to privacy or marketing concerns \cite{elaheReviewLoadData2021}.

To identify the dataset gap filled by the \textit{FeederBW} dataset, we provide an overview of existing public \ac{LV} grid datasets. We exclude datasets with highly aggregated load or market data at higher system levels, such as those presented in \cite{elaheReviewLoadData2021}. Furthermore, we distinguish for public \ac{LV} grid datasets between (1) residential, (2) commercial/industrial, and (3) grid-measured datasets \cite{habenReviewLowVoltage2021}. 

The majority of these datasets consist of smart meter data collected from residential buildings (1). These datasets provide real-world load data from single households \cite{pereiraResidentialLabeledDataset2022, trivediComprehensiveDatasetElectrical2024} and are often enriched with metadata, such as information about the appliances present in the respective buildings \cite{athanasouliasPlegmaDatasetDomestic2024}.

In contrast, commercial and industrial datasets (2) include commercial buildings in addition to residential buildings \cite{zhouHighresolutionElectricPower2023, sandellDatasetNorwegianMedium2023} or focus exclusively on industrial facilities and their associated metadata \cite{bischofHIPEEnergyStatusDataSet2018}. These datasets are valuable for analyzing energy consumption patterns in non-residential settings.

Residential, commercial, and industrial datasets can include data about the grid, such as grid topology \cite{sandellDatasetNorwegianMedium2023}. However, these datasets typically do not incorporate measurements of \ac{DSO} equipment in the \ac{LV} grid. In the \textit{FeederBW} dataset, we address this gap by focusing on measurements of the \ac{LV} feeders. This is particularly significant given the scarcity of public datasets that include such measurements, making \ac{LV} feeder data strongly underrepresented.

Datasets with measurements in the \ac{LV} grid (3) are scarcely published. Often, available grid datasets are aggregations of smart meter values from households. While smart meter data from individual customers can be aggregated to the \ac{LV} feeder level, it is important to note that this data is not directly measured at the \ac{LV} feeder and is therefore not considered as \ac{LV} grid-measured dataset in this paper. For example, such datasets typically do not include phase imbalances or a mix of residential and commercial customers. The approach of aggregating data from various consumers and producers at the smart meter level is exemplified in benchmark datasets such as SimBench \cite{meineckeSimBenchBenchmarkDataset2020}. Similarly, projects and datasets like the data from Hvaler in Norway \cite{dang-haLocalShortTerm2017}, Weave Smart Meter Data \cite{WeaveSmartMeter}, the Iowa Distribution test system project, the SSEN Distribution \ac{LV} Feeder Usage data, and UK Power Networks data provide smart meter consumption data aggregated at the \ac{LV} feeder level \cite{habenReviewLowVoltage2021}.

Unlike grid-measured datasets, grid datasets often include a grid topology, which is essential for applications such as \ac{PF} analysis, \ac{SE}, and other grid-related studies \cite{altamimiSmartGridPublic2024}. Existing \ac{LV} grid datasets serve various purposes, for example providing benchmark grids and test systems like the IEEE and CIGRE test cases \cite{peyghamiStandardTestSystems2019}. However, these datasets frequently lack time-series load data or are limited to specific grid configurations \cite{enganReferenceDatasetSemiurban2025}. As a result, researchers often rely on semi-synthetic data or other real-world load time-series for grid calculations.

In contrast, the \textit{FeederBW} dataset provides 200 \ac{LV} feeders with one-minute temporal resolution, offering a large volume of load and generation data over two years. This makes it particularly valuable for \ac{ML} applications. We include recommended train-test splits for \ac{ML} experiments in the \textit{Usage Notes} at the end of the paper. While the \textit{FeederBW} dataset does not include a grid topology, as it is not designed for grid calculations, it can supplement such calculations by providing real-world time-series data.

\Cref{tab:overview} provides an overview of \ac{LV} grid datasets that are closely related to our dataset. We have included datasets that contain time-series power or load measurements, either from \ac{MV} to \ac{LV} substations or from \ac{LV} feeders. The small number of five datasets already underlines the necessity for more published grid-measured datasets. 

\begin{table}
    \centering
    \scriptsize
    \rowcolors{2}{white}{white}
    \renewcommand{\arraystretch}{1.5}
    \begin{adjustbox}{margin*=0cm 0cm 0cm 0cm}
        \begin{tabularx}{15.6cm}{p{2cm}p{1.1cm}p{1.2cm}p{1.5cm}p{1.4cm}p{2cm}p{1.4cm}p{1.5cm}}
            \toprule
            \textbf{Name} & \textbf{Country} & \textbf{Period (Years)} & \textbf{Equipment of DSO} & \textbf{Resolution} & \textbf{Measurements} & \textbf{Weather data} & \textbf{Area type (Location)}\\
            \midrule
            BigDEAL \cite{shuklaBigDEALChallenge20222024} & US & 2015 - 2018 (4y) & 3 sub. & 1h & load & T & unclear (USA)\\
            OPSCI TP Gradisce \cite{elektroljubljanaOPSCI} & SI & 2019 - 2023 (4y) & 1 sub. & 15 min & P, Q & not given & unclear (Slovenia)\\
            NTVV \cite{scottishandsouthernenergypowerdistributionNTVVSubstationsDataset2012} & UK & 2014 - 2017 (4y) & 316 sub. including feeders & 5s & V, I, P, Q, VHC & not given & suburban (Bracknell)\\
            Flexible Networks for a Low Carbon Future \cite{spenergynetworksFlexibleNetworksLow} & UK & unclear (1y) & 184 sub. & 30 min & unclear & not given & unclear (St Andrews, Wrexham, Whitchurch\\
            Rolle Hierarchical Benchmark \cite{lorenzoHierarchicalDemandForecasting2019} & CH & 2018 (1y) & 24 sub. \& cabinets & 10 min, (weather: 1h, 12h)  & for each phase: P, Q, V, THD, $\omega$ & T, GHI, GNI, RH, Pr, $W_S$, $W_{dir}$ & rural (Rolle) \\
            \midrule
            \textbf{FeederBW} [this paper] & DE & 2023 - 2025 (2y) & 200 feeders & 1 min (weather: 1h) & for each phase: I, PF; V, P, Q & DirR, DifR, T, Prec, SH, Hum, WG, MW, ZW & mainly rural (Baden-Württem-berg)\\
            \bottomrule
        \end{tabularx}
    \end{adjustbox}
    \caption{Literature and datasets of \ac{LV} grid-measured datasets with real measurement time-series of power or load for \ac{MV}/\ac{LV} substations (sub.), \ac{LV} feeders or cable distribution cabinets. Abbreviations measurements: V (voltage), I (current), P (active power), Q (reactive power), VHC (voltage harmonic content), THD (total harmonic distortion), $\omega$ (voltage frequency), PF (power factor). Abbreviations weather: T (Temperature), GHI (global horizontal irradiance), GNI (global normal irradiance), RH (relative humidity), Pr (pressure), $W_S$ (wind speed), $W_{dir}$ (wind direction), DirR (direct radiation), DifR (diffuse radiation), Prec (precipitation), SH (snow height), Hum (humidity), WG (max wind gust), MW (meridional wind), ZW (zonal wind). Country codes: US (United States of America), UK (United Kingdom), SI (Slovenia), CH (Switzerland), DE (Germany).}
    \label{tab:overview}
\end{table}
The available datasets in \Cref{tab:overview} have several limitations that the FeederBW dataset addresses. There is a lack of \ac{LV} feeder data specific to Germany, a gap filled by FeederBW with data from Baden-Württemberg. In comparison to other countries, the power at the \ac{LV} feeder is characterized by a high degree of \ac{PV} generation \cite{hossainSolarPVHighpenetration2024}. Temporally, FeederBW provides more recent data from 2023 to 2025, aligning with the 1-4 year scope of the other datasets (2012 - 2023). Existing datasets often include only a few substations or \ac{LV} feeders, limiting their use for \ac{ML} \cite{shuklaBigDEALChallenge20222024, elektroljubljanaOPSCI}. FeederBW offers data from $200$ \ac{LV} feeders over two years, including the diversity of \ac{LV} feeders for a complete region and supporting robust training and evaluation of models.

Minute-level resolution is rare in \Cref{tab:overview}, with only \cite{scottishandsouthernenergypowerdistributionNTVVSubstationsDataset2012} providing a higher resolution of 5 seconds. FeederBW provides with minute-level resolution the second highest resolution. Additionally, while many datasets focus solely on load measurements \cite{shuklaBigDEALChallenge20222024}, FeederBW includes comprehensive data including currents, power, power factor and voltage. Lastly, weather data is often lacking \cite{elektroljubljanaOPSCI, scottishandsouthernenergypowerdistributionNTVVSubstationsDataset2012, spenergynetworksFlexibleNetworksLow} or scarce \cite{shuklaBigDEALChallenge20222024} in existing datasets. FeederBW includes extensive local weather data, crucial for accurate predictions given the increasing dependency of load and generation on weather conditions.

Most notably, \textit{FeederBW} includes extensive metadata about customers and their devices, such as heat pumps, electric vehicles or the number of housing units in the \ac{LV} grid. Such details are largely absent in other datasets, limiting the interpretability. While some datasets in \Cref{tab:overview} provide measurements from the underlying or overlying grid \cite{scottishandsouthernenergypowerdistributionNTVVSubstationsDataset2012, lorenzoHierarchicalDemandForecasting2019} or include grid topology information \cite{elektroljubljanaOPSCI}, none match the depth of consumer and producer details offered by \textit{FeederBW}. Additionally, \textit{FeederBW} addresses the lack of representation of \ac{PV} generation in existing datasets. Many datasets primarily focus on load data, despite the growing dominance of \ac{PV} in \ac{LV} grids. \textit{FeederBW} includes numerous feeders with significant \ac{PV} feed-in, providing a more comprehensive view of modern energy systems. Lastly, \textit{FeederBW} explicitly includes measurement data from \ac{LV} feeders characterized by industry and commerce, a feature often missing or unclear in other datasets.

The FeederBW dataset supports a variety of applications and research in energy systems. It can be used to develop and evaluate diverse \ac{ML} models, including load forecasting \cite{habenReviewLowVoltage2021, altamimiSmartGridPublic2024}, building load and generation profiles, and \ac{NILM} at the \ac{LV} feeder level. For example, a similar dataset was used in \cite{treutleinGeneratingPeakawarePseudomeasurements2025} to train a model for pseudo-measurements. The model uses feeder metadata, weather data, calendar data and the timestamp encoding to estimate the load and generation of non-measured \ac{LV} feeders. The active power measurements are used as target variable. 

Additionally, the dataset enables the creation of synthetic data for grid studies, allowing for more realistic simulations. Researchers can also work on data improvement techniques, such as data imputation, anomaly detection, and correction of metadata errors. Furthermore, the dataset facilitates analysis and understanding of grid dynamics, including the coincidence of devices/households at the \ac{LV} feeder level, weather influences on load and generation, and relationships between power factors, active and reactive power, and phase currents. 

Many existing datasets, as shown in \Cref{tab:overview}, suffer from poor curation, including missing background information, inaccessibility (e.g., \cite{elektroljubljanaOPSCI}, access failed on 24 September 2025), and lack of identifiers like \acp{DOI}. To ensure the quality and usability of our datasets, we adhere to the \textit{FAIR principles} \cite{wilkinsonFAIRGuidingPrinciples2016}. We use Zenodo, which assigns \acp{DOI} to each entry, to enable findability. Zenodo also provides multiple retrieval methods and ensures long-term accessibility. The download of our dataset is manageable on conventional computers. We employ standard data types and file formats for interoperability and provide extensive metadata and licensing information to clarify reusability and provenance.

\section*{Methods}\label{sec2} 

\begin{figure}
    \centering
    \includegraphics[width=1\linewidth]{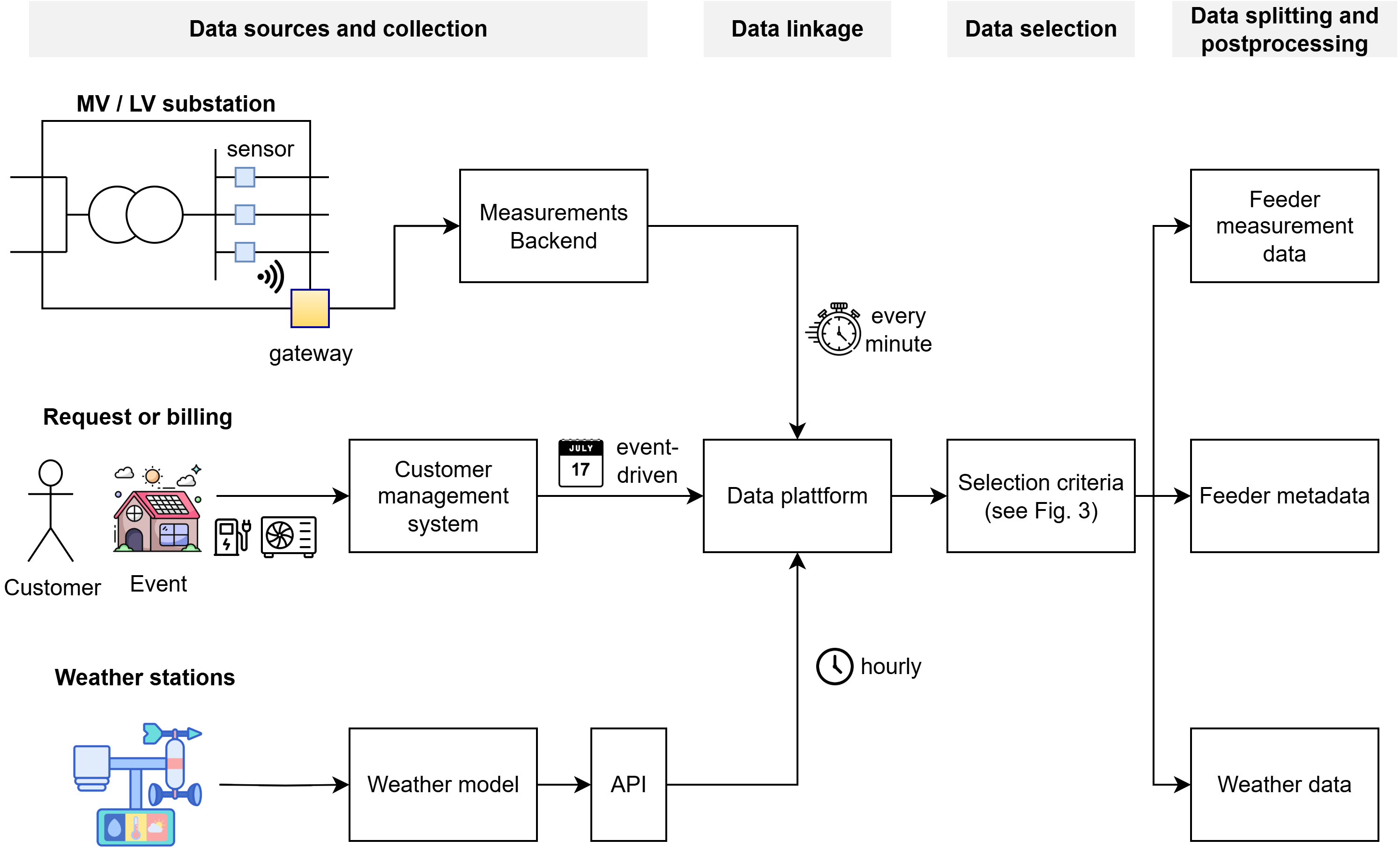}
    \caption{The data collection process of the feeder, meta and weather data. The collected data is combined in a data platform, the data is selected and again split into three data records. [icons in image: flaticon.com]}
    \label{fig:data_process}
\end{figure}

The published data is real-world data collected and used at Netze BW, the largest \ac{DSO} in the federal state of Baden-Württemberg in the southwest of Germany. We describe the collection of the feeder data which is enriched by metadata and weather data on a data platform. Furthermore, we describe which criteria were applied to derive the final $200$ \ac{LV} feeders. The data process is shown in \Cref{fig:data_process}.

\subsection*{Data sources and collection}\label{subsec:data_sources_and_collection}

\paragraph{Feeder measurement data}

\begin{figure}
    \centering
    \includegraphics[width=0.9\linewidth]{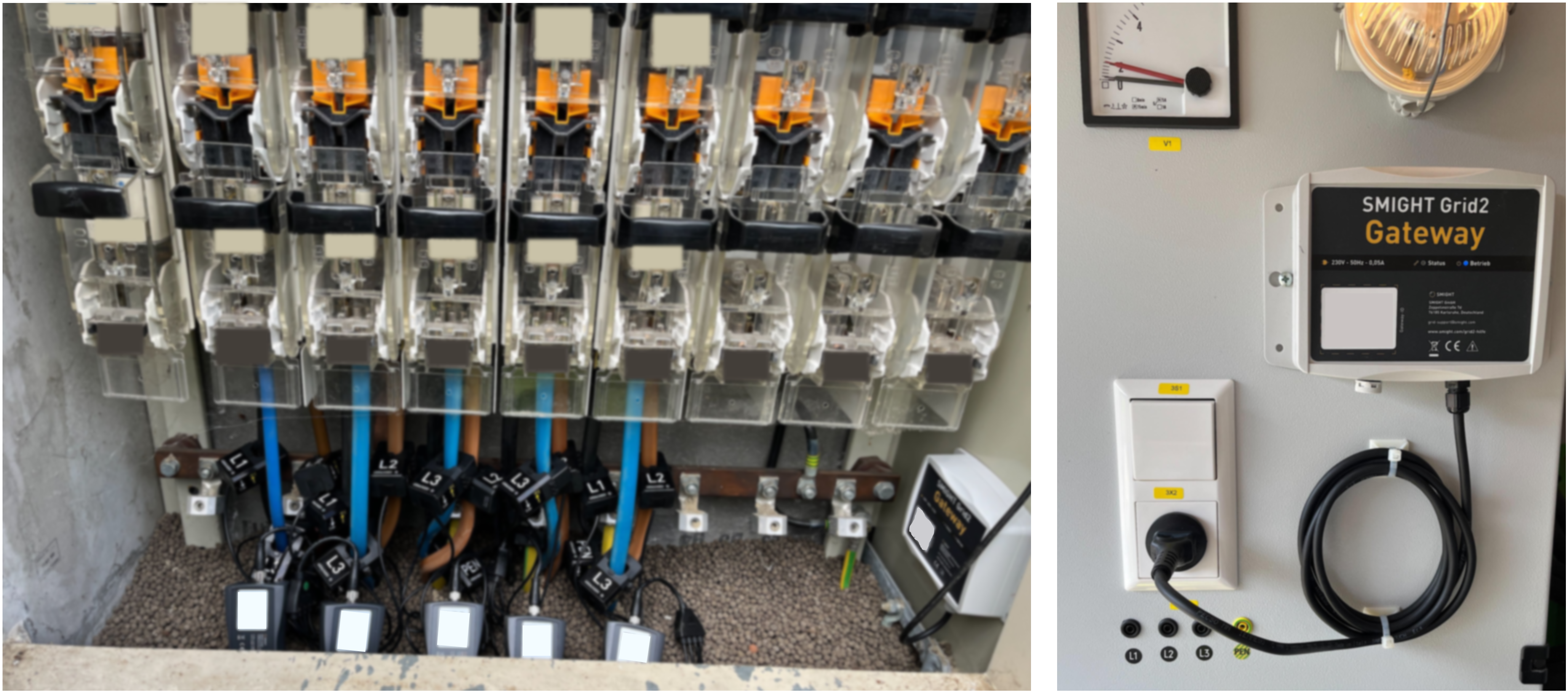}
    \caption{Example for sensors and gateway used in the \ac{MV}/\ac{LV} substations to measure the \ac{LV} feeders. (image: Netze BW GmbH)}
    \label{fig:smight_illustration}
\end{figure}

The measurement of the feeders is conducted using \textit{SMIGHT Grid 2 (gateway + sensor)} \cite{smightBetriebsanleitungSMIGHTGrid2022} from the company SMIGHT, installed in \ac{MV}/\ac{LV} substations (see \Cref{fig:smight_illustration}). These sensors are clipped around each phase of a \ac{LV} feeder cable. The transformer is not directly measured. The sensors capture the \ac{RMS} current for each phase, including \ac{PEN} for each minute. Power supply for the sensors is achieved through energy harvesting, eliminating the need for an external power source. However, this necessitates a minimum current for the sum of all phases plus \ac{PEN} of $32$~\si{\ampere} as energy supply for a stable operation of the sensor. The measurement accuracy is $3$~\% between $5$ - $400$~\si{\ampere}, resulting in $\pm 11.85$~\si{\A}. 

The collected data is transmitted to a gateway. It measures the busbar voltage synchronized to the current measurement by directly capturing the voltage of its own energy supply. Since the energy supply of the gateway is provided by only one phase, we only measure the voltage on one phase. The measurement range is between $85$ - $264$~\si{\volt} with $3$~\% accuracy. Phase angle, the direction of the current flow and the distribution between active and reactive part of the current are determined based on the phase shift between current and voltage. Finally, the gateway forwards the data to the backend. It is important to note that there is only one voltage measurement for the entire substation but not for individual feeders or phases. Consequently, feeders from the same substation share identical voltage measurements.

\paragraph{Feeder metadata}

The feeder metadata originates from the customer management system at Netze BW and can be categorized into three types: the number of connected objects, the installed power of consumers or producers, and the average hourly energy consumption of industry and commerce customers. Note that the industry values are averages over several months or years and not hourly data. The data collection process spanned several decades and involved numerous employees across various regions, leading to slight variations in the methodology in addition to changes in the data acquisition process over time. Unlike feeder measurement data and weather data, this dataset is characterized by event-driven entries, triggered by customer connection requests such as new \ac{PV} installations. Furthermore, we update the entry when new average energy consumption values for industry and commerce are recorded. Finally, the metadata of a \ac{LV} feeder given in this dataset describes the sum of all registered connected objects, the sum of installed power for consumers and producers and the current average energy consumption for industry and commerce.

\paragraph{Weather data}

The weather data is retrieved from an API based on the zip code where the substation of the \ac{LV} feeder is located. The API is caching weather forecasts with a forecast horizon of one to three hours and averages the meteorological variables of a zip code area. The weather data itself is provided by the \ac{NWP} model ICON-D2 from the German Meteorological Service (DWD)~\cite{reinertDWDDatabaseReference2024}, which relies on data from several hundred weather stations across Germany and offers spatially resolved weather forecasts. The weather forecasts have an hourly resolution.

\subsection*{Data linkage}

The linkage of the data was conducted on a data platform introduced in \cite{bottcherProfessionellesDatenmanagementForm2024}. We combined customer data, such as the number of installed \ac{PV} systems or \ac{EV} chargers, with measurements based on the grid topology to obtain the feeder metadata. In the subsequent step we retained only non-meshed grids to ensure that the feeder metadata could be clearly assigned to a single \ac{LV} feeder rather than multiple feeders. The \ac{LV} feeder is situated in an area identified by a zip code, which was used to match weather data with feeder measurement data and feeder metadata.

\subsection*{Data selection process}\label{subsec:data_selection_process}

\begin{figure}
    \centering
    \includegraphics[width=0.9\linewidth]{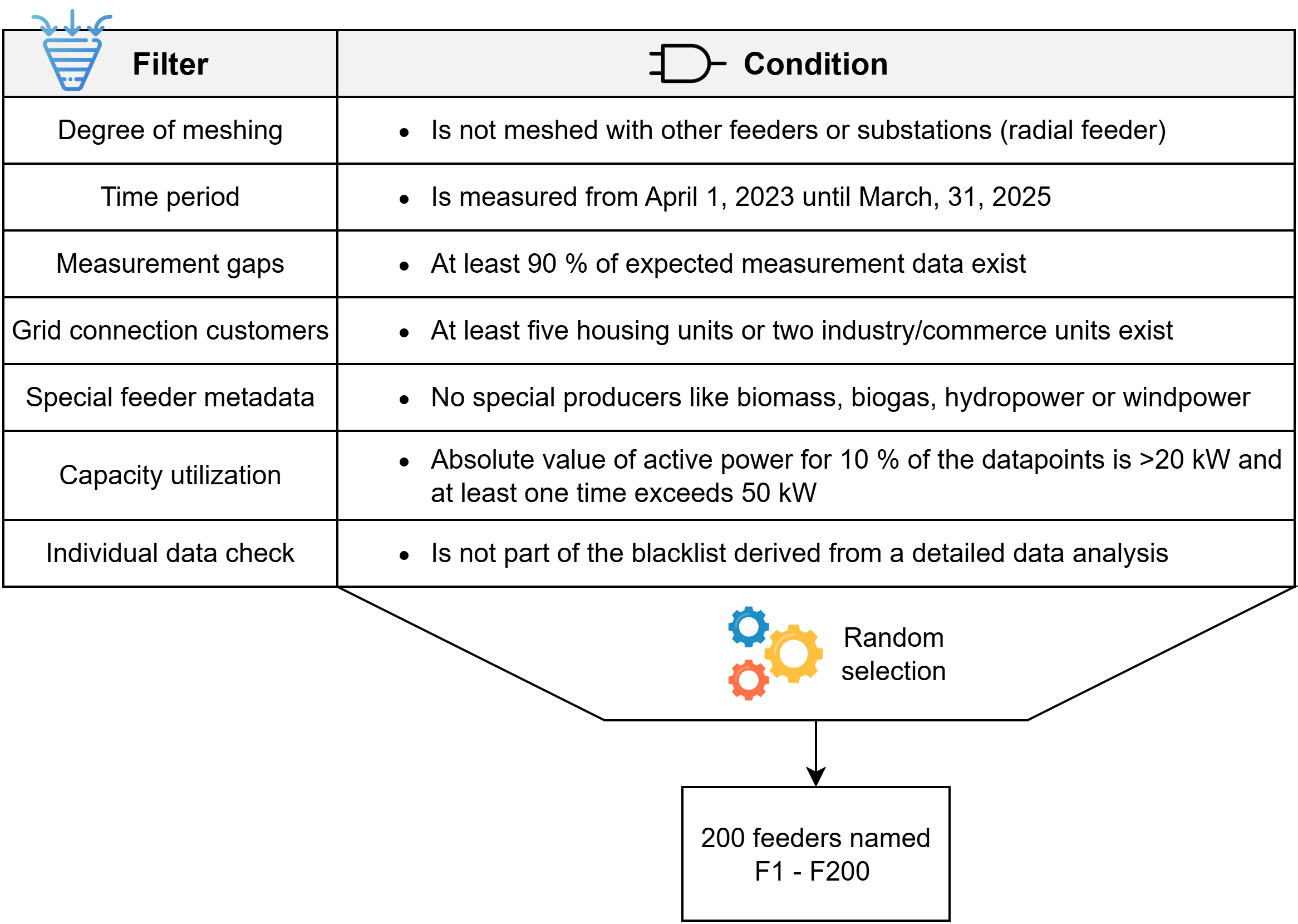}
    \caption{Selection criteria for the $200$ \ac{LV} feeders published in the FeederBW dataset. All the published feeders fulfill all conditions. Refer to \Cref{fig:data_process} for the complete data process where the data selection is embedded. [icons in image: flaticon.com]}
    \label{fig:filter_process}
\end{figure}

Every \ac{LV} feeder included in the \textit{FeederBW} dataset must satisfy all conditions outlined in \Cref{fig:filter_process}.

\textbf{Degree of meshing:} This filter ensures the feeder is radial, preventing interconnections with other feeders or substations. Thereby we ensure that the load and feed-in of producers and consumers given in the feeder metadata can be fully assigned to the feeder without splitting the power flow across multiple feeders.

\textbf{Time period:} This filter ensures measurements span from April 1, 2023, to March 31, 2025, providing a two-year window which allows meaningful train-test splits for \ac{ML} applications (see \textit{Usage Notes}).

\textbf{Measurement gaps:} This filter guarantees that at least $90$~\% of expected measurement data is available, reducing errors and minimizing preprocessing for users.

\textbf{Grid connection customers:} This filter requires at least $5$ housing units or $2$ industry/commerce units to ensure data privacy, complementing the pseudonymization of feeder names.

\textbf{Special feeder metadata:} This filter excludes feeders with rare producers like biomass, biogas, hydropower, or wind power due to their unique and sparse data.

\textbf{Capacity utilization:} This filter ensures that the absolute value of the active power exceeds $20$~\si{\kW} for $10$~\% of data points and reaches at least $50$~\si{\kW} once, focusing on feeders with significant capacity utilization, which are the crucial \ac{LV} feeders for \acp{DSO} with respect to operation and planning. 

\textbf{Individual data check:} This filter uses a blacklist based on detailed analysis to exclude feeders with uncorrectable data quality issues, such as topology changes.

Finally, $200$ feeders are randomly selected for publication. While these criteria introduce bias, they ensure high data quality, privacy, information security, and relevance for \ac{LV} feeders with notable capacity utilization.

\subsection*{Data splitting and postprocessing}

For the data processing steps, such as data selection, it was convenient to have the data combined to apply filters. For publication, the data is split into feeder measurement data, feeder metadata, and weather data to reduce storage requirements due to different temporal resolutions. The resolution of the feeder measurement data is one minute, whereas the resolution of the weather data is one hour. The feeder metadata can be considered as event data, remaining constant until a new consumer or producer is registered or deregistered, or the average hourly energy consumption of industry/commerce changes.

For 51 \ac{LV} feeders, all \ac{PEN} values are in the range $1$ - $2$~\si{\A}. The same accounts for the \ac{LV} feeder $181$ in the first months. The values for these feeders are replaced with Null, because it can be assumed that the instrument transformer could not be installed properly around the \ac{PEN} cable. 

\section*{Data Records}\label{sec3}


The dataset is hosted on Zenodo \cite{zenodo_this_paper}. As illustrated in \Cref{fig:data_process}, it is divided into three distinct data records: one for feeder measurement data, one for feeder metadata, and one for weather data. These records are summarized in \Cref{tab:data_records_summary} and the columns are presented in detail in \Cref{tab:feeder_measurement_data} (feeder measurement data), \Cref{tab:feeder_metadata} (feeder metadata) and \Cref{tab:weather_data} (weather data).

Each detailed table includes for each column of the data record the name, unit, data type, and a brief description. All three data records include a \textit{feeder} column, which is an integer specifying the feeder (ranging from 1 to 200). Additionally, the feeder measurements and the weather data include \textit{timestamp} columns in both UTC and CET/CEST which is the local time of the substations. The metadata only includes a date. The unique identifier (primary key) for each data point in the datasets is a combination of the feeder plus timestamp (UTC) or date.

\begin{table}
    \centering
    \scriptsize
    \rowcolors{2}{white}{white}
    \renewcommand{\arraystretch}{1.5}
    \begin{adjustbox}{margin*=0cm 0cm 0cm 0cm}
        \begin{tabularx}{13.76cm}{p{3cm}p{3cm}p{3cm}p{3cm}}
            \toprule
            \textbf{Dataset} & \textbf{Feeder measurement data} & \textbf{Feeder metadata} & \textbf{Weather data}\\
            \midrule
            Number of columns & 20 & 22 & 12\\
            Number of rows & 209,189,649 & 1,608 & 3,508,800 \\
            Temporal resolution & 1 minute & event-driven & 1 hour\\
            File format & parquet & CSV & parquet\\
            File size & 8.5 GB (compressed) & 171 KB & 106 MB\\
            Description & Voltage, power, currents, power factor of the \ac{LV} feeder & Number of customer connection objects, installed power of consumers and producers and average consumption of industry/commerce at \ac{LV} feeder & Weather data such as radiation, temperature and wind in zip code area of \ac{LV} feeder\\
            \bottomrule
        \end{tabularx}
    \end{adjustbox}
    \caption{Overview about the three published data records. Details about the columns of each data record are given in \Cref{tab:feeder_measurement_data} (feeder measurement data), \Cref{tab:feeder_metadata} (feeder metadata) and \Cref{tab:weather_data} (weather data).}
    \label{tab:data_records_summary}
\end{table}

\subsection*{Feeder measurement data}

The feeder measurement data comprises 209,189,649 rows and $20$ columns, with the structure detailed in \Cref{tab:feeder_measurement_data}. For each minute and \ac{LV} feeder, the dataset includes the voltage of one phase, power values for the sum of all three phases, current values for each phase (including PEN), and power factors for each phase, unless data is missing for the specific feeder and timestamp. \ac{PEN} stands for protective earth neutral and describes a conductor which is used for grounding and as the neutral conductor. A timestamp, for example April 1, 2023 at 10:05 am specifies that the measurement was taken a few milliseconds before 10:05 am at that day. We choose to represent the data in the units V, A, kW, kVAr and kVA which are most suited for the voltage, current and power ranges in \ac{LV} grids. Consequently, with the exception of the power values, only the first three decimal places are relevant. The additional decimal places are present due to the floating point representation. 

Positive active power indicates a power flow from the \ac{MV} grid to the \ac{LV} grid whereas the active power is negative if the power flows from the \ac{LV} grid in the direction of the \ac{MV} grid. The reactive power is positive for inductive behavior and negative for capacitive behavior. Additionally, if the voltage is zero, a default value of $230$~\si{\V} is used to facilitate the calculation of apparent, active, and reactive power values. The data is stored in parquet format to enable efficient compression of the extensive dataset \cite{vohraApacheParquet2016}. The parquet format is column-oriented which means that the data is stored by column and not by row.

\begin{table}
    \centering
    \scriptsize
    \rowcolors{2}{white}{white}
    \renewcommand{\arraystretch}{1.5}
    \begin{adjustbox}{margin*=0cm 0cm 0cm 0cm}
        \begin{tabularx}{14.9cm}{p{4cm}p{1cm}p{1.5cm}p{6.65cm}}
            \toprule
            \textbf{Column name} & \textbf{Unit} & \textbf{Data type} & \textbf{Description}\\
            \midrule 
            feeder & count & int & Pseudonym of the feeder\\
            timestamp\_UTC & - & string & Timestamp in time zone UTC\\
            timestamp\_CET\_CEST & - & string & Timestamp in local time zone of the DSO (Central European (Summer) Time)\\
            \midrule
            \makecell[lt]{voltage\_one\_phase\_V \\ (abbr. $U$)} & V & float & Voltage of one phase measured at busbar/gateway\\
            \midrule
            active\_power\_kW & kW & float & Calculated based on the voltage $U$ of one phase and the respective active currents with $U \cdot (I_{L_{1},act} + I_{L_{2},act} + I_{L_{3},act}) \cdot \frac{1}{1000}$ \\
            reactive\_power\_kVar & kVar & float & Calculated based on the voltage $U$ of one phase and the respective reactive currents with $U \cdot (I_{L_{1},rea} + I_{L_{2},rea} + I_{L_{3},rea}) \cdot \frac{1}{1000}$ \\
            apparent\_power\_kVA & kVA & float & Calculated based on the voltage $U$ of one phase and the respective apparent currents with $U \cdot (I_{L_{1},app} + I_{L_{2},app} + I_{L_{3},app}) \cdot \frac{1}{1000}$ \\
            \midrule
            \makecell[lt]{l1\_active\_current\_A \\ (abbr. $I_{L_{1},act}$)} & A & float & Active current of $L_1$ computed on gateway based on \acf{RMS} current measured at sensor and the phase shift with respect to the voltage $U$ of one phase measured at gateway\\
            \makecell[lt]{l1\_reactive\_current\_A \\ (abbr. $I_{L_{1},rea}$)} & A & float & Reactive current of $L_1$ computed on gateway based on \acf{RMS} current measured at sensor and the phase shift with respect to the voltage $U$ of one phase measured at gateway\\
            \makecell[lt]{l1\_apparent\_current\_A \\ (abbr. $I_{L_{1},app}$)} & A & float & Apparent \ac{RMS} current of phase $L_1$ measured at the sensor\\
            l1\_power\_factor\_cos\_phi & - & float & Power factor of phase $L_1$ based on phase shift between current and voltage\\
            \midrule
            \makecell[lt]{l2\_active\_current\_A \\ (abbr. $I_{L_{2},act}$)} & A & float & Analog to phase $L_1$\\
            \makecell[lt]{l2\_reactive\_current\_A \\ (abbr. $I_{L_{2},rea}$)} & A & float & Analog to phase $L_1$\\
            \makecell[lt]{l2\_apparent\_current\_A \\ (abbr. $I_{L_{2},app}$)} & A & float & Analog to phase $L_1$\\
            l2\_power\_factor\_cos\_phi & - & float & Analog to phase $L_1$\\
            \midrule
            \makecell[lt]{l3\_active\_current\_A \\ (abbr. $I_{L_{3},act}$)} & A & float & Analog to phase $L_1$\\
            \makecell[lt]{l3\_reactive\_current\_A \\ (abbr. $I_{L_{3},rea}$)} & A & float & Analog to phase $L_1$\\
            \makecell[lt]{l3\_apparent\_current\_A \\ (abbr. $I_{L_{3},app}$)} & A & float & Analog to phase $L_1$\\
            l3\_power\_factor\_cos\_phi & - & float & Analog to phase $L_1$\\
            \midrule
            lpen\_apparent\_current\_A & A & float & Apparent \ac{RMS} current of \ac{PEN} conductor measured at the sensor\\
            \bottomrule
        \end{tabularx}
    \end{adjustbox}
    \caption{Description for columns of the data record feeder measurement data. The data is recorded at a resolution of one minute, and each row is uniquely identified by the feeder and the timestamp (UTC). It contains voltage, power, currents for each phase including \ac{PEN} and the power factor for each phase.}
    \label{tab:feeder_measurement_data}
\end{table}

\subsection*{Feeder metadata}

\Cref{tab:feeder_metadata} outlines the columns of the feeder metadata, where each of the $22$ columns (except \textit{feeder} and \textit{date}) represent metadata attributes of the feeder, $20$ in total. Due to the relative stability of feeder metadata compared to dynamic feeder measurements and weather data, the data record contains only 1,608 rows. Each row specifies the metadata of a \ac{LV} feeder for the time interval starting at the row's date and ending at the date of the next chronological row for that feeder.

The feeder metadata includes the number of housing units and industry and commerce in the underlying \ac{LV} grid of the feeder. Furthermore, it includes the installed power of consumers such as various heating systems, \ac{EV} chargers, public lighting and other consumers. The installed power of batteries can be both load and feed-in. Feed-in is provided by \ac{PV} and \ac{CHP} systems.

To enhance information regarding industry and commerce, the dataset includes aggregated average hourly energy consumption data for seven different customer types. It is important to clarify that these values represent averages over long periods (months or years) rather than a time-series of hourly consumption data. The values are updated irregularly because the consumption data originates from several industrial and commercial customers with different measurement settings. The six different customer groups of industry and commerce plus the general category $g_0$ follow the classification of a publication about representative load profiles \cite{vdewReprasentativeVDEWLastprofile1999}. However, this only applies to the classification of customer groups, the data itself has nothing to do with representative load profiles. Customer type $g_1$ is for load during workdays from $8$ a.m. to $6$ p.m. while $g_2$ is for industrial/commercial customers which predominantly consume in the evening hours. Customer type $g_3$ is from buildings with a constant load demand. Shops are categorized into $g_4$ while the bakeries get an extra category with $g_5$. The category $g_6$ is industry and commerce with its main consumption on the weekend. Finally, $g_0$ is for all industry and commerce not falling into $g_1$ - $g_6$.

The dataset is stored in CSV format to ensure compatibility with programs that may not support compressed parquet files, facilitating quick and easy access to the data.

\begin{table}
    \centering
    \scriptsize
    \rowcolors{2}{white}{white}
    \renewcommand{\arraystretch}{1.5}
    \begin{adjustbox}{margin*=0cm 0cm 0cm 0cm}
        \begin{tabularx}{14.9cm}{p{4.2cm}p{1cm}p{1.5cm}p{6.45cm}}
            \toprule
            \textbf{Column name} & \textbf{Unit} & \textbf{Data type} & \textbf{Description}\\
            \midrule 
            feeder & count & int & Pseudonymization of the feeder\\
            date & - & string & Metadata in the row is valid from this date until the next row of the \ac{LV} feeder\\
            \midrule 
            housing\_units\_count & count & int & Number of housing units (houses or apartments) \\ 
            industry\_and\_commerce\_count & count & int & Number of industry and commerce \\ 
            \midrule
            storage\_heaters\_kW & kW & float & Installed power of storage heaters \\ 
            heat\_pumps\_kW & kW & float & Installed power of heat pumps \\ 
            electric\_heaters\_kW & kW & float & Installed power of electric heaters  \\ 
            hot\_water\_tanks\_kW & kW & float & Installed power of hot water tanks \\ 
            flow-type\_heaters\_kW & kW & float & Installed power of flow-type heaters \\ 
            EV\_chargers\_kW & kW & float & Installed power of \ac{EV} chargers \\ 
            public\_lighting\_kW & kW & float & Installed power of public lighting\\ 
            other\_consumers\_kW & kW & float & Installed power of other consumers, e.g. inductive loads\\ 
            \midrule
            batteries\_kW & kW & float & Installed power of batteries\\ 
            \midrule
            PV\_systems\_kW & kW & float & Installed power of \ac{PV} systems\\ 
            chp\_kW & kW & float & Installed power of \acp{CHP}\\ 
            \midrule
            g0\_general\_kWh & kWh & float & Hourly average consumption of commerce and industry over a longer period of time from type general\\ 
            g1\_workdays\_kWh & kWh & float & Hourly average consumption of commerce and industry over a longer period of time from type workdays\\ 
            g2\_evening\_kWh & kWh & float & Hourly average consumption of commerce and industry over a longer period of time from type evening\\ 
            g3\_continuous\_kWh & kWh & float & Hourly average consumption of commerce and industry over a longer period of time from type continuous\\ 
            g4\_shop\_hairdresser\_kWh & kWh & float & Hourly average consumption of commerce and industry over a longer period of time from type shop/hairdresser\\ 
            g5\_bakery\_kWh & kWh & float & Hourly average consumption of commerce and industry over a longer period of time from type bakery\\ 
            g6\_weekend\_kWh & kWh & float & Hourly average consumption of commerce and industry over a longer period of time from type weekend\\
            \toprule
        \end{tabularx}
    \end{adjustbox}
    \caption{Description for columns of the data record feeder metadata. A row specifies the \ac{LV} feeder metadata from the date until the next chronological row for that feeder. Each row is uniquely identified by the feeder and the date. The data comprises the number of housing units as well as industry and commerce, installed power of equipment and aggregated industrial consumption data.}
    \label{tab:feeder_metadata}
\end{table}

\subsection*{Weather data}

The weather data described in \Cref{tab:weather_data} corresponds to individual feeders and their respective timestamps. The weather data for feeder $i$ is specific to the zip code where the \ac{LV} feeder $i$ is located. The dataset comprises 3,508,800 rows and $12$ columns, including direct and diffuse radiation, air temperature, precipitation, snow height, humidity, maximum wind gust, meridional wind, and zonal wind.  Global radiation is the sum of direct and diffuse radiation. Meridional wind is the wind speed in north-to-south or south-to-north direction and zonal wind represents the wind speed in east-to-west or west-to-east direction. The column \textit{Processing} describes if the value is an average for this hour, for one point in time (instantaneous) within this hour, accumulated throughout the hour or the maximum of the hour. For example, \textit{direct\_radiation\_W/m2} with timestamp April 1, 2023 at 10 am is the average direct radiation between 10 and 11 am.  

\begin{table}
    \centering
    \scriptsize
    \rowcolors{2}{white}{white}
    \renewcommand{\arraystretch}{1.5}
    \begin{adjustbox}{margin*=0cm 0cm 0cm 0cm}
        \begin{tabularx}{14.9cm}{p{4.2cm}p{1cm}p{1.5cm}p{2.0cm}p{4.01cm}}
            \toprule
            \textbf{Column name} & \textbf{Unit} & \textbf{Data type} & \textbf{Processing} & \textbf{Description}\\
            \midrule 
            feeder & count & int & - & Pseudonymization of the feeder\\
            timestamp\_UTC & - & string & - & Timestamp in time zone UTC\\
            timestamp\_CET\_CEST & - & string & - & Timestamp in local time zone of the DSO (Central European (Summer) Time)\\
            \midrule
            direct\_radiation\_W/m2 & $\frac{W}{m^2}$ & float & average & Hourly average of the downward solar direct radiation flux at the surface \\
            diffuse\_radiation\_W/m2 & $\frac{W}{m^2}$ & float & average & Hourly average of the downward solar diffuse radiation flux at the surface \\
            air\_temperature\_C & $^\circ\mathrm{C}$ & float & instantaneous & Temperature at 2 meters above ground\\
            precipitation\_kg/m2 & $\frac{kg}{m^2}$ & float & accumulated & Hourly total precipitation such as rain or snow\\
            snow\_height\_m & $m$ & float & instantaneous & Snow depth\\
            humidity\_kg/kg & $\frac{kg}{kg}$ & float & instantaneous & surface specific humidity\\
            max\_wind\_gust\_m/s & $\frac{m}{s}$ & float & maximum & Maximum wind gust at 10m above ground, collected over hourly intervals\\
            meridional\_wind\_m/s & $\frac{m}{s}$ & float & instantaneous & Meridional wind at 10m above ground\\
            zonal\_wind\_m/s & $\frac{m}{s}$ & float & instantaneous & Zonal wind at 10m above ground\\
            \bottomrule
        \end{tabularx}
    \end{adjustbox}
    \caption{Description for columns of the data record weather data. The data is provided at a resolution of one hour, and each row is uniquely identified by the feeder and the timestamp (UTC). The data covers solar radiation, temperature, precipitation, snow, humidity and wind.}
    \label{tab:weather_data}
\end{table}

\section*{Technical Validation}\label{sec4}

In this section we provide more information about why the data is plausible. We focus on the feeder measurements and the feeder metadata. In particular, we analyze active power and related feeder metadata. Active power is often used in applications such as \ac{LV} load forecasting and it is based on several measurements recorded in \Cref{tab:feeder_measurement_data}. However, we also implemented checks for the other measurement values which are the basis for the sections about missing data and limitations of feeder measurement data and feeder metadata. Furthermore, we give background information and describe known limitations in the data. 

\begin{table}
    \centering
    \scriptsize
    \rowcolors{2}{white}{white}
    \renewcommand{\arraystretch}{1.5}
    \begin{adjustbox}{margin*=0cm 0cm 0cm 0cm}
        \begin{tabularx}{15.6cm}{p{0.8cm}p{1cm}p{1cm}p{1cm}p{1cm}p{1cm}p{1cm}p{1cm}p{1cm}p{1cm}p{1cm}}
            \toprule
            \textbf{LV feeder} & \textbf{hous.} (count) & \textbf{indus.} (count) & \textbf{sto. h.} (\si{\kW}) & \textbf{heat p.} (\si{\kW}) & \textbf{e. hea.} (\si{\kW}) & \textbf{hot w.} (\si{\kW}) & \textbf{EV} (\si{\kW}) & \textbf{batt.} (\si{\kW}) & \textbf{PV} (\si{\kW}) & \textbf{g} (kWh)\\
            \midrule
            $F_{37}$ & 102 & 0 & 0 & 0 & 0 & 0 & 0 & 0 & 0 & 0\\
            $F_{160}$ & 47 - 48 & 10 & 78 & 0 & 0 & 15 & 0 & 5 - 9 & 6 - 14 & 4\\
            $F_{97}$ & 7 - 12 & 0 & 2 - 11 & 11 - 25 & 8 & 9 - 18 & 33 - 66 & 18 - 28 & 62 - 80 & 0\\
            $F_{33}$ & 0 & 2 & 0 & 0 & 0 & 0 & 0 - 44 & 0 & 77 & 8 - 15\\
            \midrule
            $F_{132}$ & 47 & 0 & 15 & 12 & 0 & 11 & 11 - 22 & 0 - 16 & 79 - 103 & 0\\
            $F_{63}$ & 13 - 15 & 5 - 6 & 68 & 0 & 11 & 0 & 11 & 0 - 11 & 6 - 78 & 0\\
            \bottomrule
        \end{tabularx}
    \end{adjustbox}
    \caption{Minimum and Maximum of the feeder metadata of the six \ac{LV} feeders which are visualized in the \textit{Technical Validation}. For every \ac{LV} feeder, the number of units, the installed power and the average consumption by industry and commerce is given. Values are rounded to the nearest whole number. Consumption of industry/commerce is summarized in one column $g$. Metadata which is zero for all six \ac{LV} feeders is not shown. Abbreviations: hous. (housing\_units), indus. (industry\_and\_commerce), sto. h. (storage\_heaters), heat p. (heat\_pumps), e. hea. (electric\_heaters), hot w. (hot\_water\_tanks), EV (EV\_chargers), batt. (batteries), PV (PV\_systems).}
    \label{tab:featured_feeders_infos}
\end{table}


\subsection*{Typical patterns in weekly profiles}\label{subsec:periodicity_in_weekly_active_power_profiles}

\begin{figure}
    \centering
    \begin{subfigure}{\linewidth}
        \centering
        \includegraphics[width=1\linewidth]{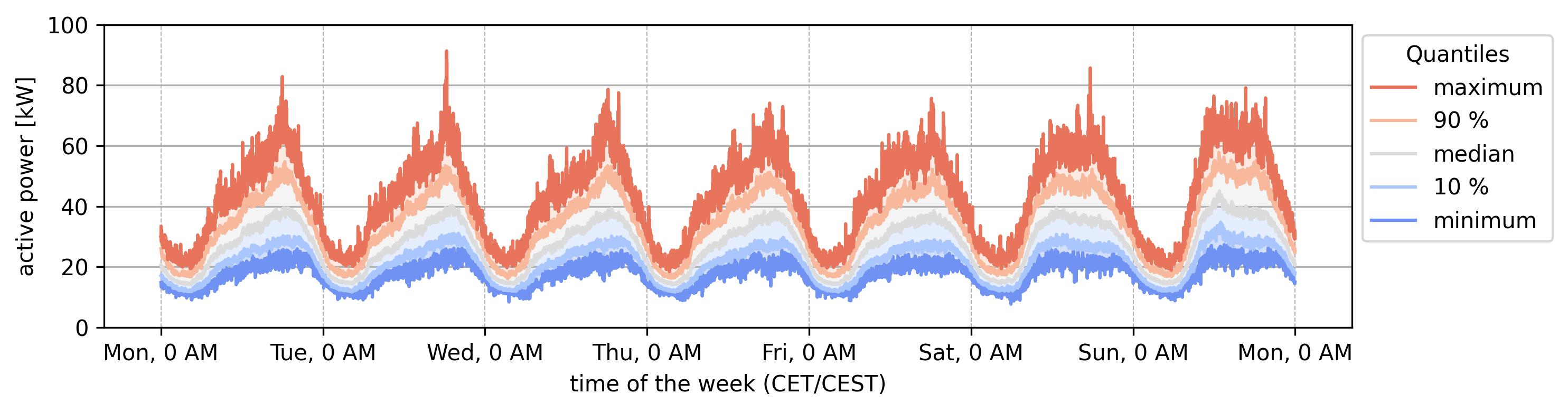}
        \caption{Feeder $F_{37}$ with $102$ housing units and no other feeder metadata.} \label{subfig:weeklyquantile_f37}
    \end{subfigure}
    \begin{subfigure}{\linewidth}
        \centering
        \includegraphics[width=1\linewidth]{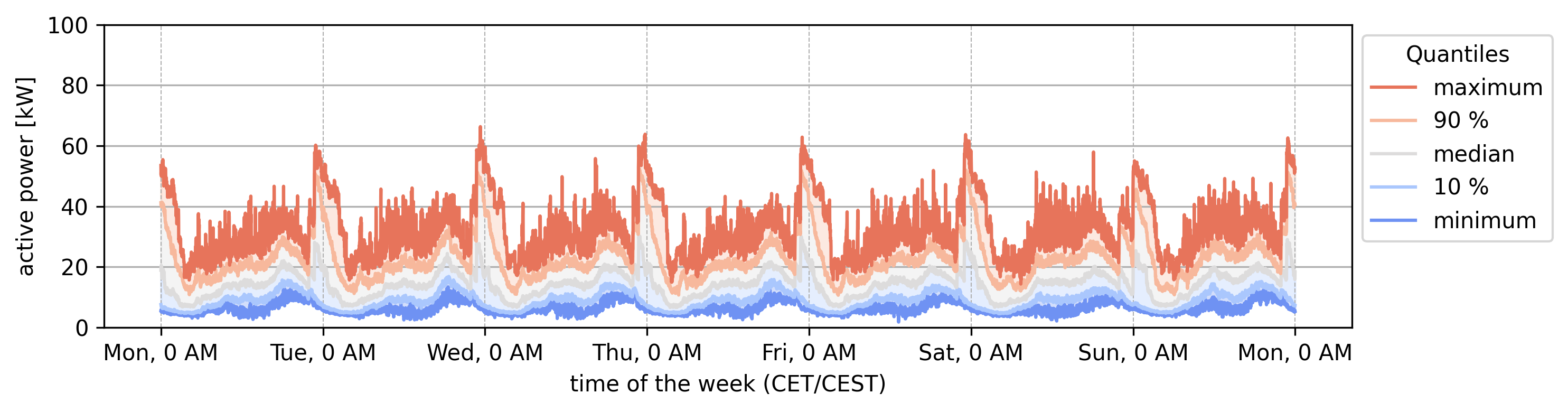}
        \caption{Feeder $F_{160}$ with $78$~\si{\kW} installed power of storage heaters are visible in the high quantiles during the night.}
        \label{subfig:weeklyquantile_f160}
    \end{subfigure}
    \begin{subfigure}{\linewidth}
        \centering
        \includegraphics[width=1\linewidth]{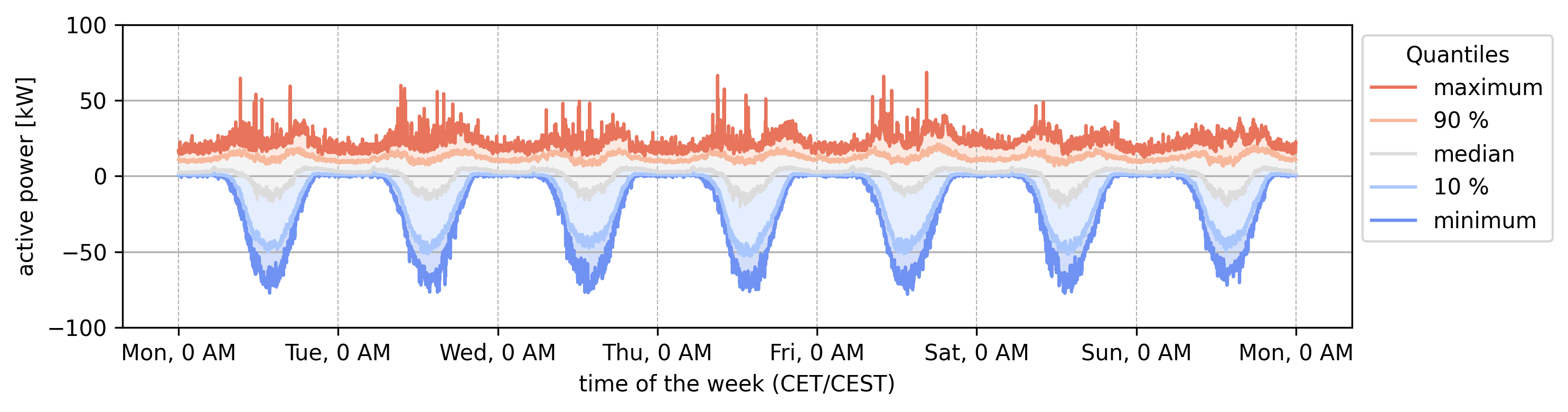}
        \caption{Feeder $F_{97}$ with $62-80$~\si{\kW} installed power of \ac{PV} systems within the two years.}
        \label{subfig:weeklyquantile_f97}
    \end{subfigure}
    \begin{subfigure}{\linewidth}
        \centering
        \includegraphics[width=1\linewidth]{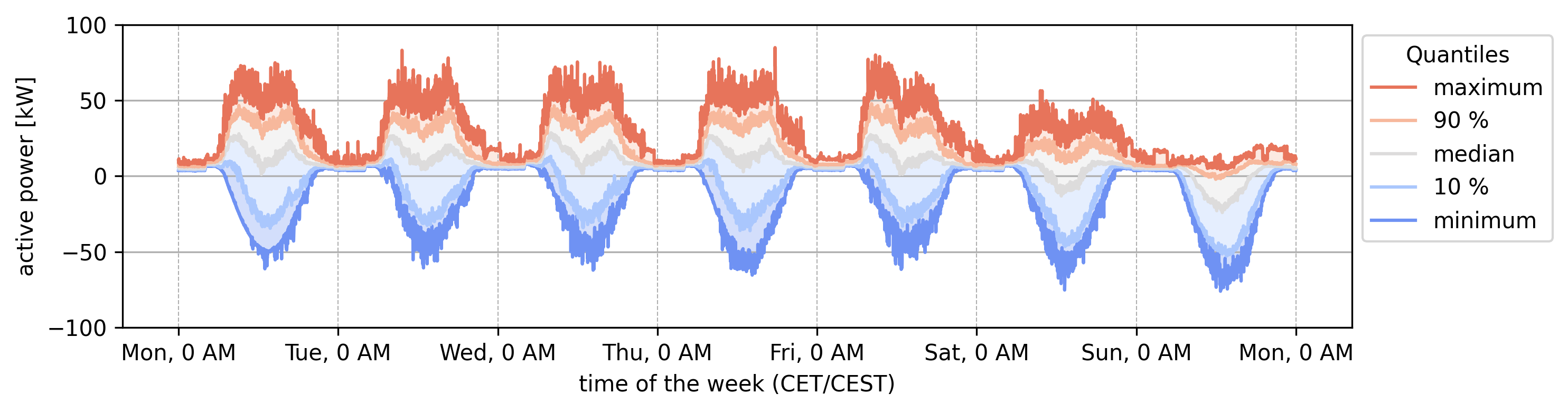}
        \caption{Feeder $F_{33}$ with two industry and commerce and $76.8$~\si{\kW} \ac{PV} systems.}
        \label{subfig:weeklyquantile_f33}
    \end{subfigure}
    \caption{Active power weekly quantile profiles from four \ac{LV} feeders showing diverse characteristics. Constructed by shrinking the two-years measurements to one week and taking the quantiles for each minute of the week. The metadata of the \ac{LV} feeders is in \Cref{tab:featured_feeders_infos}.}
    \label{fig:act_weekly_quantile_profiles}
\end{figure}

Figure~\ref{fig:act_weekly_quantile_profiles} presents aggregated weekly quantile profiles for four different \ac{LV} feeders of active power. This involves condensing the two-year active power time-series for each feeder into a single week, assigning approximately $104$ to $105$ data points to each minute of the week, and then calculating the quantiles. The periodic nature facilitates visual validation through aggregated weekly data. The \ac{LV} feeders exemplify four distinct patterns, evident in their power profile due to unique periodicity. In \Cref{tab:featured_feeders_infos} we provide the corresponding metadata information. The metadata itself already indicates that the \ac{LV} feeders are characterized by housing units ($F_{37}$), storage heaters ($F_{160}$), \ac{PV} systems ($F_{97}$) and industry and commerce ($F_{33}$).

\Cref{subfig:weeklyquantile_f37} illustrates the  \ac{LV} feeder $F_{37}$, characterized only by $102$ housing units. The load increases from morning to evening and decreases overnight, with a notable baseload as the lowest values remain significantly above 0~\si{\kW}. Additionally, the load at the weekend is higher in the late morning and at noon compared to weekdays, possibly due to more people being at home.

\Cref{subfig:weeklyquantile_f160} depicts the weekly active power profile of \ac{LV} feeder $F_{160}$, which connects, among other things, $47-48$ housing units, $10$ industry and commerce units, and 78~\si{\kW} of storage heaters to the grid. The impact of these storage heaters is evident near midnight, with sharp increases and decreases in load in the median and higher quantiles. This is expected, because storage heaters are activated synchronously. 

In contrast to $F_{37}$ and $F_{160}$, the \ac{LV} feeder $F_{97}$ in \Cref{subfig:weeklyquantile_f97} includes a significant installed power of \ac{PV} systems with $62-80$~\si{\kW} over the two years. Accordingly, we observe feed-in from the \ac{LV} feeder to the \ac{MV} grid, as indicated by the minimum crossing $-50$~\si{\kW}.

\Cref{subfig:weeklyquantile_f33} presents feeder $F_{33}$ with two units of the category industry and commerce and no housing units at all. Increased load during typical working hours and a reduced demand for load during the weekend can be observed. Furthermore, feed-in from the \ac{PV} systems with $77$~\si{\kW} is visible in the low quantiles, with even lower values during the weekend.

We do not go into detail regarding quantile plots for other columns in \Cref{tab:feeder_measurement_data}. Nevertheless, we note that periodic behavior is also observed in several other columns, such as reactive power with periodicity linked to industrial activity or \ac{PV} system feed-in. Since both motors and pumps in industry or inverters of \ac{PV} systems have an influence on reactive power, this is expected.  

\subsection*{Seasonal effects on power values}

\begin{figure}
    \centering
    \begin{subfigure}{\linewidth}
        \centering
        \includegraphics[width=1\linewidth]{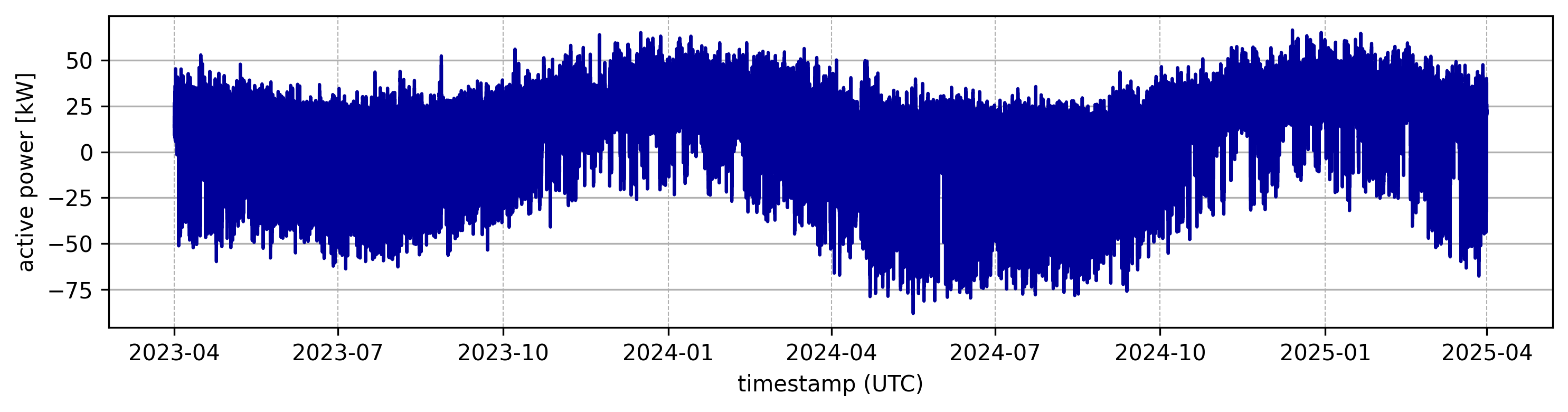}
        \caption{Feeder $F_{132}$ with $47$ housing units showing a yearly periodicity.} 
        \label{subfig:plausibility_f132}
    \end{subfigure}
    \begin{subfigure}{\linewidth}
        \centering
        \includegraphics[width=1\linewidth]{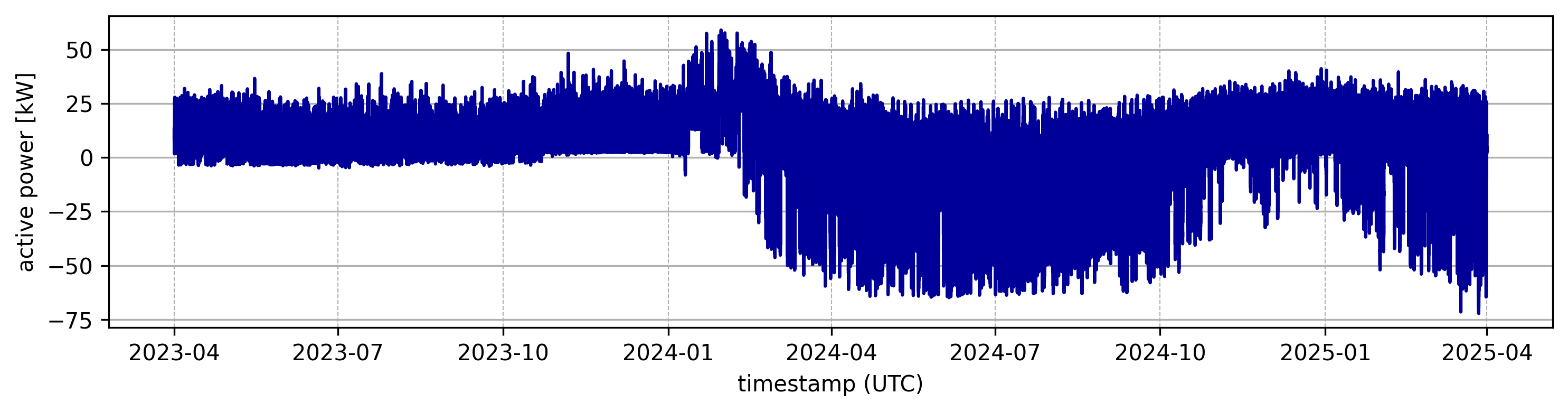}
        \caption{Feeder $F_{63}$ with a significant installation of $59.16$~\si{\kW} of \ac{PV} systems at January 10, 2024.}
        \label{subfig:plausibility_f63}
    \end{subfigure}
    \caption{Active power plotted over two years for selected \ac{LV} feeders. A seasonal effect in winter and summer is visible through higher loads and higher feed-in.}
    \label{fig:whole_time_series_plot}
\end{figure}

\Cref{subfig:plausibility_f132} illustrates the complete two-year active power values for feeder $F_{132}$, revealing a yearly periodicity with increased feed-in during summer and higher load in winter. \Cref{subfig:plausibility_f63} highlights a concept drift in the active power of feeder $F_{63}$ over two years, attributable to the installation of \ac{PV} systems at January 10, 2024, which significantly increased feed-in. This drift allows for a comparative analysis of data before and after the \ac{PV} installation. An increase of the maximum feed-in in the second year compared to the first year is also visible for $F_{132}$.

\subsection*{Two-years increase of equipment installations}

\begin{figure}
    \centering
    \includegraphics[width=1\linewidth]{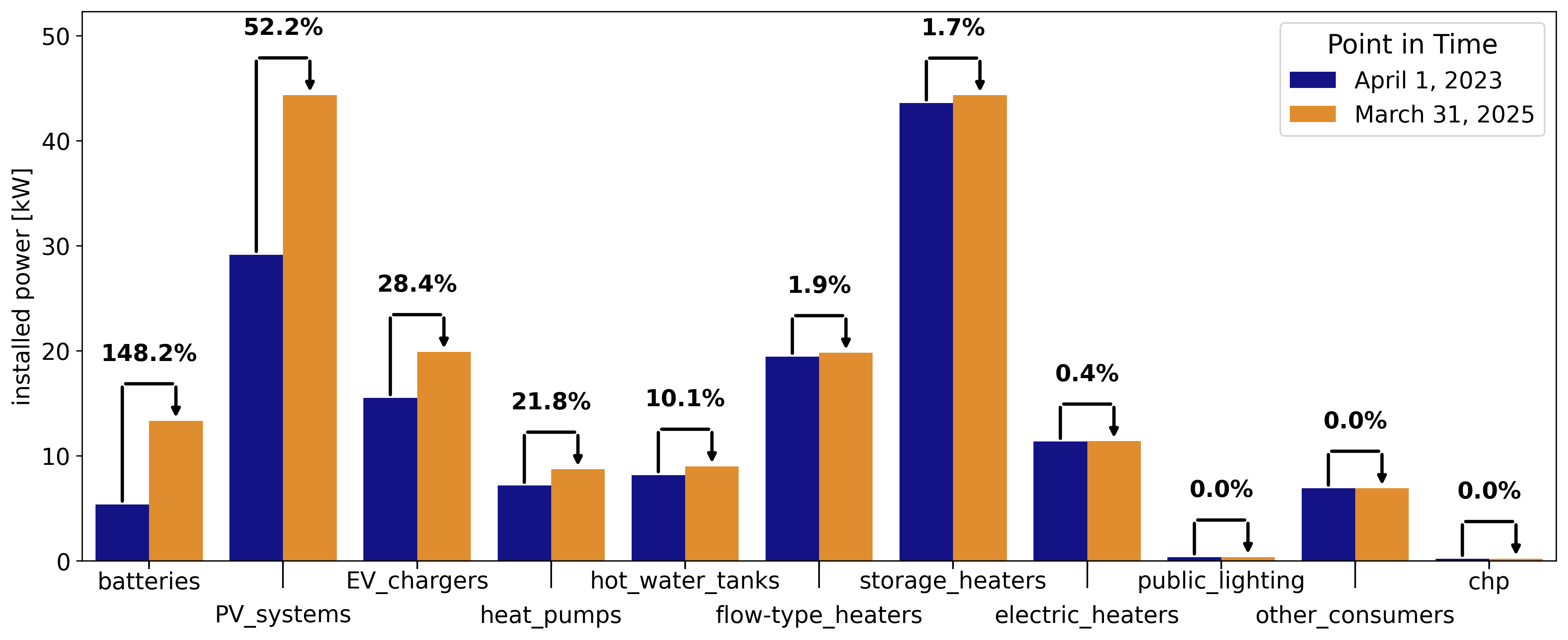}
    \caption{Average installed power per \ac{LV} feeder of all consumers and producers in the feeder metadata from April 1, 2023 to March 31, 2025. The increase of typical \acp{LCT} characterizing the energy transition is clearly visible. The '\_kw' suffix of each column name was removed.}
    \label{fig:metadata_change}
\end{figure}

In \Cref{fig:metadata_change}, we observe the change in the mean installed power per \ac{LV} feeder for consumers and producers from the first day of measurement data in 2023 to the last day in 2025. A clear increase in installed power is evident for specific equipment, notably batteries, \ac{PV} systems, \ac{EV} chargers, and heat pumps, which are key drivers of the energy transition. Batteries exhibit the highest increase of $148.2$~\%, driven by widespread installations in Germany due to falling prices and government support programs \cite{figgenerDevelopmentBatteryStorage2022}. The installed power of \ac{PV} systems rises by $52.5$~\% across the $200$ feeders, starting from an average of $29.08$~\si{\kW} per \ac{LV} feeder \cite{burgerEnergycharts2025}. \ac{EV} chargers and heat pumps also show significant increases of $28.4$~\% and $21.8$~\%, respectively, though these were less pronounced compared to batteries and \ac{PV} systems, aligning with lower-than-expected sales of new \acp{EV} and heat pumps \cite{schillAmpelMonitorEnergiewendeAmbitionierte2024}. For other metadata only hot water tanks exhibited a notable increase of $10.1$~\%, with all other changes remaining below $2$~\%.

\subsection*{Limitations in the feeder measurement data}

Even though we only visualize active power values in the technical validation (see \Cref{fig:act_weekly_quantile_profiles} and \ref{fig:whole_time_series_plot}), the feeder measurement data comprises many more electrotechnical variables as depicted in \Cref{tab:feeder_measurement_data}. We conducted extensive checks and validation to all of them to ensure a good quality of data and be able to describe limitations. 

The dataset comprises one-minute \ac{LV} feeder measurements from $200$ \ac{LV} feeders, with an expected total of 210,528,000 measurements. The expected number of rows results from $200$ \ac{LV} feeders measured over two years whereas one year has $365.5$ days, $24$ hours and $60$ minutes. The feeder measurement data record contains 209,189,649 rows, indicating a completeness of $99.36$~\% and only $0.64$~\% missing values. Regarding missing values in existing rows (NULL values), there are only $833$ measurements of voltage missing ($0.0004$~\%) and $25.6$~\% of \ac{PEN} values. As described in the methods, $51$ \ac{LV} feeders have no \ac{PEN} values and $F_{181}$ has only around two third because they were removed in a postprocessing step. In most cases, the reason is that the instrument transformer could not be installed to measure the \ac{PEN} due to limited installation space in the substation.  

We note that the power values are approximate, particularly in scenarios with low current and harmonics, where deviations may occur. This is not only due to the $3$~\% accuracy of the current and voltage. The voltage is measured only on one phase which can lead to deviations when determining the power factors of the individual phases. 

Furthermore, it can be observed in the data that the equation $S = \sqrt{P^2 + Q^2}$ for apparent power $S$, active power $P$ and reactive power $Q$ is not fulfilled. However, the large deviations can be explained with opposite current and power flows on the phases and represent valid measurements. The largest deviation when calculating $dev = |S - \sqrt{P^2 + Q^2}|$ for all data points is $dev = 38.06$~kVA for \ac{LV} feeder $F_{127}$ on April 9, 2023 13:49 CEST. The power factors $\lambda$ are $\lambda_{L_1} = -1$, $\lambda_{L_2} = -1$ and $\lambda_{L_3} = 1$, indicating opposite power flows of the phases. If the data points are filtered based on the total power factor $\lambda_{tot} = \sum \lambda_{L_{1}} + \lambda_{L_{2}} + \lambda_{L_{3}}$ with $\lambda_{tot} < -2$ or $\lambda_{tot} > 2$, the highest deviation is only $dev = 9.46$~kVA for $F_{9}$ on November 3, 2023 10:27 CET with $\lambda_{tot} = 2.09$.

Opposite flows on the phases can be caused by single-phase inverters of \ac{PV} systems. The choice between using active or apparent power in your experiments depends on your research question. The active power indicates a net load or net feed-in of the underlying \ac{LV} grid. In contrast to the apparent power, the active power also includes a direction of the power flow. However, during times of high reactive power flows or opposite power flows on the phases, the active power is insufficient to indicate the thermal utilization of the cable or overhead line. The example of \ac{LV} feeder $F_{127}$ on April 9, 2023 13:49 CEST illustrates this. Here, the active currents $I_{L_{1},act} = -43$~\si{\A}, $I_{L_{2},act} = -48$~\si{\A} and $I_{L_{1},act} = 84$~\si{\A} almost cancel each other out and the reactive currents are between $-2$~\si{\A} and $1$~\si{\A}. This results in a low active power of $-1.7$~\si{\kW} and a low reactive power of $-0.5$~kVar while at the same time the apparent power is $39.9$~kVA. Hence, the formula $S = \sqrt{P^2 + Q^2}$ is not applicable here.

Regarding the power factor, we observe that the values quickly change within a few minutes if the direction of the power flow changes. This is often the case shortly after sunrise and shortly before sunset when the \ac{PV} feed-in increases or levels off. The direction flips the power factor to the opposite, because the power factor $\lambda$ is defined by $\lambda = \frac{P}{S}$.  Furthermore, the current is very low and therefore out of the effective range of the sensor during the change of the power flow direction. Consequently, the inversion of the power flow direction and the decreased accuracy due to low currents lead to the fluctuations of power factors.

For some \ac{LV} feeders, outliers are present in the measurement data, such as unusually high active or reactive power values resulting from high current measurements. The causes of these outliers are diverse and often difficult to identify or correct, as they may reflect valid measurements. Therefore, we did not do any postprocessing regarding outliers. Only one current value is greater than the effective measurement range of $400$~\si{\A} ($F_{134}$ on August 12, 2023 15:01 CEST). The same applies to the upper bound of the voltage, which has one value exceeding the effective measurement range of $264$~\si{\V} ($F_{16}$ and $F_{88}$ from the same substation on July 5, 2023 10:24 CEST). Regarding the lower bound of the voltage measurement range, $71$ data points are below $85$~\si{\V}. Apart from two consecutive data points of $F_{145}$, all data points belong to $F_{38}$, $F_{94}$ and $F_{187}$. These three \ac{LV} feeders show a significant voltage drop for $77$ minutes in the period November 4, 2024 11:47 CET to November 4, 2024 13:04 CET. It is unclear if this is a valid measurement. 

\subsection*{Limitations in the feeder metadata}\label{subsec:issues_in_the_metadata}

The feeder metadata can be considered as time-series event data and therefore we do not have missing timestamps analog to the feeder measurements. We carefully conduct the assignment of metadata to a \ac{LV} feeder. However, we point out that the assignment can be inaccurate for some \ac{LV} feeders due to technical restrictions. Furthermore, we note that equipment in the \ac{LV} grid can be unregistered. This implicates that existing feeder metadata can be missing or wrong in the data record. Furthermore, the day of registration in the feeder metadata and the day for the start of operation can differ, for example when \ac{PV} systems are registered but not yet active. 

During operation over the two-year period, switching events in the grid altered the topology, often due to construction work. These events are not documented in the data and normally lead to concept drifts. While obvious cases were removed, some may still be present. In such instances, the feeder metadata may not accurately represent the consumers and producers, affecting the measurement data at the feeder sensor. In contrast, natural concept drifts, such as changes in weather or customer behavior, also occur. Concept drifts can additionally manifest in reactive power, influencing apparent power.

Storage heaters, a technology primarily installed decades ago, remain active and are observable in measurements due to their periodic activation during winter nights. However, the metadata may not accurately reflect the power of these storage heaters if they are not deregistered with the \ac{DSO}. Automatic removal of incorrect storage heater data is challenging due to varying activation times and usage intensity. Furthermore, missing information about already deregistered equipment affects also the other consumers and producers. 

Regarding the different customer groups of industry and commerce, it should be noted that even though the distinction in the dataset is sharp, this is not necessary the case for individual industrial and commercial customers. For example, a bakery is assigned to group bakery ($g_5$). However, while this is meaningful for a traditional bakery which bakes bread locally and at night, bakeries only selling bread may be more similar to shops and hairdressers ($g_4$) for shops and hairdresser. Another example are restaurants which can have their main electricity usage in the evening ($g_2$) or in the weekend ($g_6$).

\subsection*{Limitations in the weather data}
The weather data record does not exhibit missing rows in the dataset, so $100$~\% of the data is existing. Null values only occur in column \textit{snow\_height\_m}, primarily in the summer and autumn months from June to October in the year 2024. In applications and models you can assume $0$~\si{\m} because there is usually no snow at this season.

\subsection*{Further background information}\label{subsec:further_background_information}

We shortly summarize some aspects in the context of the dataset that influence the interpretation of the measured data. The capacity of a \ac{LV} feeder depends on several factors, including the conductor size and the installation. Therefore, it is not possible to derive grid congestions or capacity limits from the data. For example, the ampacity of overhead lines is higher compared to underground cables for the same conductor size \cite{deleonMajorFactorsAffecting2006}.

The data is collected after accounting for grid losses in the \ac{LV} grid, which are mostly below $1.3$~\% with a median of $0.45$~\% according to comparable literature \cite{maEvaluationEnergyLosses2019}. Consequently, the influence of network topology and the distance between equipment and measurement at the \ac{LV} feeder on the measured values is moderate.

Regarding price elasticity and demand response, it is expected to be not or only slightly observable in the measurement data. Even though there is a obligation since 2025 to offer dynamic tariffs to customers and dynamic network charges are present for some customers, a wide usage is not given, in particular due to missing smart meter installations \cite{kueblerEreignisvariableTarifeZur2024}. Some customers are still prized based on a high (often hours during the day) and low tariff (often hours during the night) which was introduced some decades ago to make better use of base load power plants \cite{heuckElektrischeEnergieversorgungErzeugung2013}. Some customers may also use special tariffs for large consumers such as heat pumps.

It is important to clarify that the installed power of \ac{EV} chargers does not directly correspond to the presence of \acp{EV} in the grid. Funding programs for private \ac{EV} charging stations have led to the installation of chargers in anticipation of future \acp{EV}.

The dataset does not differentiate between single-family and multi-family houses. However, most \ac{LV} feeders are located in rural areas, predominantly featuring single-family houses. A higher number of housing units indicates the presence of multi-family houses.

The method of room heating significantly influences the load on the \ac{LV} feeders. A higher degree of electrification for heating increases the load. During 2023–2025, the majority of houses still rely on fossil fuels such as gas or oil for heating~\cite{statistaHeizungsmarktDeutschland2024}. Gas is commonly used in buildings connected to gas pipelines.

Batteries in the dataset are primarily used to store energy from \ac{PV} systems and supply it directly to the house. Some batteries also have the capability to draw power from the grid or feed it back, depending on their configuration.

\section*{Usage notes}\label{sec:usage_notes}


The data can be accessed in Zenodo. The data records have a total size of $8.5$~\si{\giga\byte} (compressed) and are provided in the format parquet and CSV. The feeder metadata and the weather data is provided in one file. The feeder measurement data is contained in $200$ files (one file for each \ac{LV} feeder) and compressed into five zip files with $40$ parquet files respectively. The data can be loaded with common python libraries such as polars, pandas or other tools.

We encourage scientists conducting \ac{ML} experiments with the dataset to follow a consistent split of the data into training and test to enhance comparability. We recognize that deviations may be appropriate depending on the research question or methodology. For locational splits, we recommend to use feeders $F_{1} - F_{160}$ for training and $F_{161} - F_{200}$ for testing. When conducting a cross validation, we recommend using a five-fold-cross validation with the folds $F_{1} - F_{40}$ (test feeders in the first iteration), $F_{41} - F_{80}$ (test feeders in the second iteration), $F_{81} - F_{120}$ (test feeders in the third iteration), $F_{121} - F_{160}$ (test feeders in the fourth iteration) and $F_{161} - F_{200}$ (test feeders in the fifth iteration). The locational splits correspond to a random distribution. 

For temporal splits, we recommend to use the first year (April 1, 2023 to March 31, 2024) for training and the second year (April 1, 2024 to March 31, 2025) for testing. In addition, we encourage users to apply cross validation for time-series forecasting. We do not recommend folds for the time-series cross validation because they can differ based on your research question.

\section*{Data availability}

The data is available under Zenodo \cite{zenodo_this_paper} with the link: \href{https://zenodo.org/records/17831177}{https://zenodo.org/records/17831177}. The repository contains the data records of the feeder measurement data, the feeder metadata and the weather data. This arXiv preprint belongs to the Zenodo dataset version v1.0.

\section*{Code availability}

There is no code available.

\section*{Acknowledgements}
We thank Netze BW for providing the data in corporation with their partners at SMIGHT and EnBW Datalab.

\section*{Author contributions}

M.T., P.B. and M.S. designed the dataset and implemented code for the dataset extraction. M.T. and R.M. conceptualized the paper. M.T., P.B. and O.N. analyzed the data. M.T. wrote the first draft. P.B., M.S., R.H., O.N., R.M. and V.H. proofread the paper, added explanations and improved readability.

\section*{Competing interests}

The authors declare no competing interests.

\section*{Funding}
A funding was provided by the Helmholtz Association through the 'Energy System Design' program and the Helmholtz Association's Initiative Helmholtz AI. Open Access funding enabled and organized by Project DEAL.

\end{document}